\documentclass[prd,showpacs,preprintnumbers,amsmath,amssymb,superscriptaddress,nofootinbib]{revtex4}
 
\usepackage{graphicx}
\usepackage{dcolumn}
\usepackage{bm}
\usepackage{color}
\usepackage{caption}

\def\be{\begin{equation}}
\def\ee{\end{equation}}
\def\bea{\begin{eqnarray}}
\def\eea{\end{eqnarray}}

\allowdisplaybreaks[2]

\begin{document}


\title{Three-loop QCD corrections to the decay constant of $B_c$}

\author{Feng Feng\footnote{f.feng@outlook.com}}
\affiliation{Institute of High Energy Physics and Theoretical Physics Center for Science Facilities, Chinese Academy of Sciences, Beijing 100049, China\vspace{0.2cm}}
\affiliation{China University of Mining and Technology, Beijing 100083, China\vspace{0.2cm}}

\author{Yu Jia\footnote{jiay@ihep.ac.cn}}
\affiliation{Institute of High Energy Physics and Theoretical Physics Center for Science Facilities, Chinese Academy of Sciences, Beijing 100049, China\vspace{0.2cm}}
\affiliation{School of Physics, University of Chinese Academy of Sciences, Beijing 100049, China\vspace{0.2cm}}

\author{Zhewen Mo\footnote{mozw@ihep.ac.cn }}
\affiliation{Institute of High Energy Physics and Theoretical Physics Center for Science Facilities, Chinese Academy of Sciences, Beijing 100049, China\vspace{0.2cm}}
\affiliation{School of Physics, University of Chinese Academy of Sciences, Beijing 100049, China\vspace{0.2cm}}

\author{Jichen Pan\footnote{panjichen@ihep.ac.cn }}
\affiliation{Institute of High Energy Physics and Theoretical Physics Center for Science Facilities, Chinese Academy of Sciences, Beijing 100049, China\vspace{0.2cm}}
\affiliation{School of Physics, University of Chinese Academy of Sciences, Beijing 100049, China\vspace{0.2cm}}
\author{Wen-Long Sang~\footnote{wlsang@swu.edu.cn}}
\affiliation{School of Physical Science and Technology, Southwest University, Chongqing 400700, China\vspace{0.2cm}}

\author{Jia-Yue Zhang\footnote{zhangjiayue@ihep.ac.cn}}
\affiliation{Institute of High Energy Physics and Theoretical Physics Center for Science Facilities, Chinese Academy of Sciences, Beijing 100049, China\vspace{0.2cm}}
\affiliation{School of Physics, University of Chinese Academy of Sciences, Beijing 100049, China\vspace{0.2cm}}

\date{\today}

\begin{abstract}
Within the framework of nonrelativistic QCD (NRQCD) factorization,
we compute the three-loop QCD corrections to the decay constant of $B_c$.
We reconstruct the analytical expressions for the three-loop renormalization constant and the corresponding anomalous dimension affiliated with
the pseudoscalar current composed of two different heavy flavors in NRQCD, which are functions of the ratio between the charm and bottom quark masses.
Meanwhile, the short-distance coefficient is obtained with very high numerical accuracy. The three-loop QCD correction turns out to be overwhelmingly
large. The phenomenological implication of this new piece of radiative corrections for the $B_c$ leptonic decay is also addressed.
\end{abstract}

\maketitle

The $B_c$ meson is a unique member among the heavy quarkonia family. Since the $B_c$ consists of two different flavours of heavy quarks,
its decay is necessarily initiated by the weak interaction, consequently its lifetime $\tau_{B_c}\approx 0.51$ ps
is much longer than the familiar charmonium and bottomonium counterparts whose decay mechanisms are dominated by
the strong/electromagnetic interactions.
There has been tremendous amount of theoretical endeavors to unravel the myth beneath this special quarkonium~\cite{Chang:1992pt, Chang:2000ac, Chang:2001pm,Chen:2020ecu}.
The $B_c$ meson was first discovered by \texttt{CDF}~\cite{CDF:1998ihx}  and \texttt{D0} Collaborations~\cite{Corcoran:2005ti} at the Fermilab Tevatron
through the semi-leptonic decay $B_c \to J/ \psi + l +\nu _l$ by the end of last century.
More production and decay channels of the $B_c$ family have also been measured at \texttt{LHC} experiment~\cite{LHCb:2014ebd,LHCb:2017vlu,LHCb:2013kwl,LHCb:2021tdf,LHCb:2014glo}.
Needless to say, more precise theoretical predictions about the $B_c$ meson's
property appears to be highly desirable.

The aim of this work is to report a new progress on the accurate prediction of a basic observable concerning the $B_c$,
the decay constant of the $B_c$ meson,  $f_{B_c}$.
As a fundamental nonperturbative parameter characterizing the strength of the leptonic decay of $B_c$, the decay constant $f_{B_c}$ is defined by
the vacuum-to-$B_c$ matrix element mediated by the axial vector current:
\bea\label{eqn:dc full}
\langle 0| \bar{b} \gamma^{\mu} \gamma_5 c |B_c \rangle= i  f_{B_c} P^{\mu},
\eea
where $P^\mu$ is the 4-momentum of the $B_c$. Here the $B_c$ state in the left-hand side admits the standard relativistic normalization,
so the $f_{B_c}$ carries the unit mass dimension.

To date the leptonic decay of the $B_c$ has not yet been observed, therefore the direct experimental input for $f_{B_c}$ is lacking.
However, there have been many theoretical attempts to predict the value of $f_{B_c}$, based on various phenomenological
approaches, exemplified by the quark potential model~\cite{Song:1986ix, Ikhdair:2005xe, Ikhdair:2003ry, Ikhdair:2004hg, Fulcher:1993sk, Eichten:1995ch},
QCD sum rules~\cite{Aliev:1992vp, Onishchenko:2000yy, Baker:2013mwa, Narison:2014ska, Narison:2019tym, Narison:2020guz},
and lattice simulations~\cite{McNeile:2012qf, Colquhoun:2015oha, Colquhoun:2016osw, Becirevic:2018qlo}, and so on.

As a widely-accepted doctrine, the $B_c$ meson should be viewed as a genuine heavy quarkonium state rather than
a heavy-light meson such as the $B$, $D$ mesons, whose constitutes, the $c$ and $\bar{b}$ move non-relativistically.
Therefore, it is appropriate to employ the nonrelativistic QCD (NRQCD) effective field theory to describe the $B_c$ meson.
In accordance with the spirit of the NRQCD factorization~\cite{Bodwin:1994jh}, the fact that $m_{b,c} \gg \Lambda_{\rm QCD}$
indicates that the $B_c$ decay constant needs not to be an entirely nonperturbative quantity.
At the lowest order in velocity expansion,  the $B_c$ decay constant can be separated into a
perturbatively calculable short-distance coefficient (SDC) multiplied with
the nonperturbative yet universal NRQCD long-distance matrix element (LDME):
\bea\label{eqn:fac}
f_{B_c}= \sqrt{\frac{2}{M_{B_c}}}\, \mathcal{C}(m_b,m_c,\mu_{\Lambda})
\, \langle 0 |\chi^{\dagger}_b\psi_c(\mu_{\Lambda})|B_c \rangle+\mathcal{O}(v^{2}),
\eea
where $\chi^{\dagger}_b$ and $\psi_c$ denote the Pauli spinor fields  annihilating the $\bar{b}$ and $c$ quarks, respectively.
$\mathcal{C}$ denotes the dimensionless SDC, as a function of $m_c$, $m_b$ and the NRQCD factorization scale $\mu_{\Lambda}$.
It is worth noting that the $B_c$ state inside the NRQCD LDME is normalized in the non-relativistic convention.

During the past three decades, we have continuously witnessed the impressive progress in computing the higher-order corrections to $f_{B_c}$ in \eqref{eqn:fac}.
The order-$\alpha_s$ and order-$v^2$ corrections to $f_{B_c}$ was first calculated by Braaten and Fleming in 1995~\cite{Braaten:1995ej}.
Later on the order-$v^2$ relativistic correction, with partial high-order relativistic corrections resummed, was also investigated~\cite{Lee:2010ts}.
The two-loop radiative ${\cal O}(\alpha_s^2)$ corrections to $\mathcal{C}$  in \eqref{eqn:fac}
was first explored by Onishchenko {\it et al.} in 2003, which nevertheless attempted to present the result in an
asymptotic series in the limit $m_c \ll m_b$~\cite{Onishchenko:2003ui}. The complete analytical expression of the two-loop QCD corrections to $\mathcal{C}$
was finally achieved by Chen and Qiao in 2015~\cite{Chen:2015csa}. The two-loop QCD radiative corrections appear to be negative yet modest,
less important than the one-loop QCD radiative correction. One then naturally wonders how important the three-loop QCD corrections would be.

In passing, we note that the main results of the three-loop QCD corrections to      the $\Upsilon$($J/\psi$) decay constants have already been available about
a decade ago~\cite{Marquard:2006qi,Marquard:2009bj,Marquard:2014pea,Beneke:2014qea,Egner:2022jot}
(for a very recent refinement of the three-loop QCD corrections to $\Upsilon$($J/\psi$) leptonic width, also see \cite{Feng:2022vvk}).
In contrast, the knowledge of the three-loop QCD corrections to $f_{B_c}$ is still missing. It is conceivable that the
calculation in the $B_c$ case is technically much more demanding, since the $B_c$ involves two different mass scales
while $\Upsilon$($J/\psi$) only consists of a single mass scale.

In this work, we present the long-awaiting result for the three-loop QCD ${\cal O}(\alpha_s^3)$ correction to $f_{B_c}$ with high numerical accuracy.
We find that the effect of this new piece of the QCD radiative correction is substantial.

On general physical consideration, we find it is convenient to decompose the dimensionless SDC $\mathcal{C}$ in powers series of the strong coupling constant
in the following specific form:
\begin{align}
\label{eqn:sdc:decomposition}
& \mathcal{C}(m_b,m_c,\mu_{\Lambda},\mu_{R}) =1+\frac{\alpha_s^{\left(n_f\right)}\left(\mu_R\right)}{\pi} \mathcal{C}^{(1)}(x)+\left(\frac{\alpha_s^{\left(n_f\right)}\left(\mu_R\right)}{\pi}\right)^2
\left(\mathcal{C}^{(1)}(x)\frac{\beta_0}{4}\text{ln}\frac{\mu_{R}^2}{m_M^2}
+\gamma^{(2)}(x)\ln \frac{\mu_{\Lambda}^2}{m_M^2}+\mathcal{C}^{(2)}(x)\right)
\nonumber \\
& +\left(\frac{\alpha_s^{\left(n_f\right)}\left(\mu_R\right)}{\pi}\right)^3\left\lbrace \left(\frac{\mathcal{C}^{(1)}(x)}{16}\beta_1+\frac{\mathcal{C}^{(2)}(x)}{2}\beta_0\right)\text{ln}\frac{\mu_{R}^2}{m_M^2}
+\frac{\mathcal{C}^{(1)}(x)}{16}\beta^2_0 \ln^2
\frac{\mu_{R}^2}{m_M^2}+\frac{1}{4}\left(2\frac{d\gamma^{(3)}(x,\mu_{\Lambda})}{d \text{ln}\mu_{\Lambda}^2}-\beta_0\gamma^{(2)}(x)\right)\ln^2\frac{\mu_{\Lambda}^2}{m_M^2} \right.
\nonumber \\
& \left. +\left(\mathcal{C}^{(1)}(x) \gamma^{(2)}(x)+\gamma^{(3)}(x,m_M)\right)\ln\frac{\mu_{\Lambda}^2}{m_M^2}+\frac{\beta_{0}}{2}\gamma^{(2)}(x)\ln\frac{\mu_{\Lambda}^2}{m_M^2}\,\text{ln}\frac{\mu_{R}^2}{m_M^2}
+ \mathcal{C}^{(3)}(x) \right\rbrace+\mathcal{O}\left(\alpha_s^4\right),
\end{align}
where $T_F=1/2$, $C_F=(N_c^2-1)/(2N_c)$, $C_A=N_c$, and $N_c=3$ is the number of colors.
$\mu_R$ and $\mu_\Lambda$ refer to the QCD renormalization scale and NRQCD factorization scale, respectively.
$\beta_0=(11/3)C_A-(4/3) T_F n_f$ and $\beta_1=(34/4)C_A^2-(20/3) C_A T_F n_f-4 C_F T_F n_f$ are the one-loop and two-loop coefficients of
the QCD $\beta$ function, with $n_f$ signifying the number of active quark flavors~\footnote{Note in many work concerning higher-order QCD corrections for
quarkonium decay, $\mu_R$ has always been tacitly fixed at some specific value, say, the heavy quark mass~\cite{Czarnecki:1997vz,Beneke:1997jm,Kniehl:2006qw,Chen:2015csa}. Here we explicitly retain its dependence. Note the SDC must be $\mu_R$-independent. One readily checks from
\eqref{eqn:sdc:decomposition} that the $\cal C$ function
is renormalization-group invariant at each prescribed perturbative order.}.

To condense the notation, we have introduced several auxiliary variables in \eqref{eqn:sdc:decomposition}:
\bea\label{Def:m_M:x:z}
m_M \equiv \sqrt {m_b m_c}, \qquad  x\equiv {m_c\over m_b}, \qquad z\equiv {1\over 2}\left(x+{1\over x}\right).
\eea
where $m_M$ represents the geometric mean between $m_c$ and $m_b$, $x$ is the quark mass ratio.
Since the SDC must be symmetric under $m_b\leftrightarrow m_c$, therefore $\cal{C}$ must be invariant
under $x\leftrightarrow 1/x$.

The one-loop QCD correction to $\cal C$, denoted by $\mathcal{C}^{(1)}(x)$
in \eqref{eqn:sdc:decomposition}, assumes a particularly simple form~\cite{Braaten:1995ej}:
\begin{align}
& \mathcal{C}^{(1)}(x)=\frac{3}{4} C_F \left(\frac{x-1}{x+1}\,\ln x-2\right).
\label{eqn:sdc trivial}
\end{align}
The expression of the two-loop QCD correction $\mathcal{C}^{(2)}$ is somewhat too lengthy to be reproduced
here~\cite{Chen:2015csa}. Our key task in this work is to compute the three-loop contribution $\mathcal{C}^{(3)}$.

It is well-known that the SDC in quarkonium decay starts to develop explicit $\mu_\Lambda$ dependence at order-$\alpha_s^2$~\cite{Czarnecki:1997vz,Beneke:1997jm}.
The $\mu_\Lambda$ dependence of the SDC (or equivalently, the NRQCD LDME), is governed by the renormalization group equation in NRQCD. In our case,
the anomalous dimension affiliated with the pseudoscalar density current, denoted by $\gamma$ in \eqref{eqn:sdc:decomposition},
is defined through
\bea\label{eqn:gamma}
\gamma\left(x, {\mu^2_\Lambda\over m^2_M } \right) \equiv
{d \ln \widetilde{Z} \over d \ln \mu_\Lambda^2 } =\left(\frac{\alpha_s^{(n_l)}
\left(\mu_\Lambda\right)}{\pi}\right)^2 \gamma^{(2)}(x )
+\left(\frac{\alpha_s^{\left(n_l\right)
}\left(\mu_\Lambda\right)}{\pi}\right)^3\gamma^{(3)}\left(x, {\mu^2_\Lambda\over m^2_M }\right)+\mathcal{O}(\alpha^4_s).
\eea
Here $n_l=3$ is the number of light quarks, and $\widetilde{Z}$ denotes the renormalization constant of the NRQCD
pseudoscalar current, {\it e.g.},   $(\chi^\dagger_b\psi_c)_{\rm Bare} \equiv
{\widetilde Z}(\mu_\Lambda)  (\chi^{\dagger}_b \psi_c)_R (\mu_\Lambda)$.

The determination of the SDC is guided by the standard perturbative matching doctrine.
Replacing the nonperturbative $B_c$ state in \eqref{eqn:fac} by a free $c\bar{b}$ pair carrying
the quantum number $^1S_0^{(1)}$, one
then computes the current matrix elements in both perturbative QCD and NRQCD, and
solves for the coefficient function $\cal {C}$ order by order in $\alpha_s$.
The master formula is
\begin{align} \label{Master:formula}
 \sqrt{Z_{2,b} Z_{2,c} } \,  \Gamma_{\rm QCD} =
  \sqrt{2M_{B_c}} \, \mathcal{C}(m_b, m_c, \mu_\Lambda) \, \sqrt{\widetilde{Z}_{2,b} \widetilde{Z}_{2,c} } \,
  {\widetilde Z}^{-1}(\mu_\Lambda) \, \widetilde{\Gamma}_{\rm NRQCD} + {
  \mathcal O}(v^2).
\end{align}
The $Z_{2, Q}$ ($\widetilde{Z}_{2, Q}$) represent the heavy quark on-shell field-strength renormalization constant in QCD (NRQCD),
and $\Gamma_{\rm QCD}$ ($\widetilde{\Gamma}_{\rm NRQCD}$) denote the amputated current vertex function in QCD (NRQCD).
The renormalization constant affiliated with the axial vector current in QCD is equal to unity. In practice, one might simply neglect the relative momentum between $c$ and $\bar{b}$ prior to performing the loop integration
in the QCD side, which amounts to directly extracting SDC  from the hard loop momentum region
in the context of strategy of region~\cite{Beneke:1997zp}. Thus, at the lowest order in $v$, practically there is no
need to compute anything in the NRQCD side.
We work in Feynman gauge. Dimensional regularization (DR) with the spacetime dimensions $D=4-2\epsilon$ is
utilized throughout to regularize both UV and IR divergences.

\begin{figure}[t]
\center{
\includegraphics*[scale=1.5]{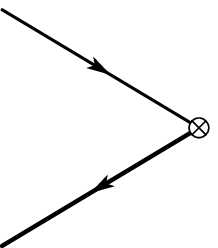}\qquad
\includegraphics*[scale=1.5]{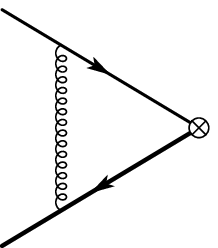}\qquad
\includegraphics*[scale=1.5]{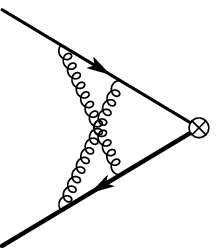}
\caption {\label{fig:bare} Representative Feynman diagrams for $c\bar{b}\to W$ through two-loop order.
The cross implies the insertion of the axial vector current.
}}
\end{figure}

\begin{figure}[t]
\center{
\includegraphics*[scale=1.5]{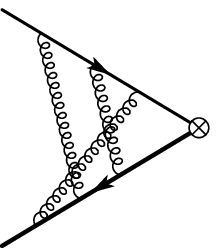}\qquad
\includegraphics*[scale=1.5]{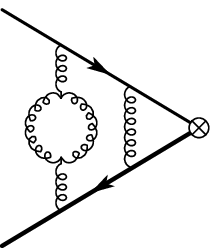}\qquad
\includegraphics*[scale=1.5]{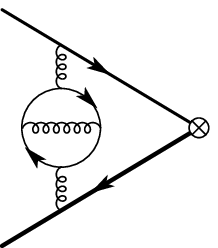}
\caption {\label{fig:bare3} Some typical Feynman diagrams for $c\bar{b}\to W$ at three-loop order.
}}
\end{figure}

To expedite the calculation, we employ the covariant spinor/color projection technique to project out the
intended QCD amplitude.
We apply the packages \texttt{QGraf}~\cite{Nogueira:1991ex} and \texttt{FeynArts}~\cite{Hahn:2000kx} to generate the corresponding Feynman diagrams and
amplitudes for $c\bar{b}({}^1S_0^{(1)})\to W$ through three-loop order in $\alpha_s$.
About 270 three-loop diagrams contribute to this process. Some representative Feynman diagrams in various perturbative order
are displayed in in Fig.~\ref{fig:bare} and Fig.~\ref{fig:bare3}.

Employing the packages \texttt{Apart}~\cite{Feng:2012iq} for partial fractions and \texttt{FIRE}~\cite{Smirnov:2014hma}
for integration-by-parts (IBP) reduction, we end up with roughly 15 master integrals (MIs) at two-loop order
and 412 master integrals at three-loop order.
Rather than use the conventional sector decomposition based packages,
We employ the newly released package~\texttt{AMFlow}~\cite{Liu:2022chg} to compute all the
multi-loop MIs. This package is based on the numerical differential equation algorithm dubbed the ``Auxiliary Mass Flow" method~\cite{Liu:2017jxz,Liu:2020kpc,Liu:2021wks},
and proves to be highly efficient to tackle MIs containing multi scales.

After implementing the quark mass and field strength on-shell renormalization, and renormalizing the QCD coupling constant
with the $\overline{\rm MS}$ prescription, the QCD amplitude is free from UV poles, yet still contains a uncancelled single
IR pole at two-loop order, and
contains some uncancelled double and single IR poles at three-loop order. As is well known, these IR
poles are intimately related to the fact
that the pseudoscalar current in NRQCD requires renormalization.
To warrant the $\cal C$ in \eqref{Master:formula} to be infrared finite, one can readjust
the renormalization factor $\widetilde{Z}$ so as to exactly cancel those residual IR poles in the QCD vertex amplitude.

Since the renormalization factor  $\widetilde{Z}$ is a function of $x$ rather than a constant,
reconstruction of its analytical form is somewhat challenging.
We have computed \eqref{Master:formula}  with several different values of the mass ratio $x$.
With the aid of the very high numerical accuracy offered by \texttt{AMFlow}, after some trial and error,
we have successfully reconstructed the exact form of $\widetilde{Z}$
by utilizing Thiele's interpolation formula~\cite{abramowitz1964handbook} and \texttt{PSLQ} algorithm~\cite{ferguson1999analysis}.
Here we just present the final result.
Through the order-$\alpha_s^3$,  the renormalization constant for the NRQCD pseudoscalar current in the $\overline{\rm MS}$ scheme
can be expressed as~\footnote{
Note here the number of active flavor in $\alpha_s$ is $n_f=n_l=3$ rather than $n_f=n_l+n_c+n_b=3+1+1=5$.
We have decoupled the effects of charm and bottom quarks in the QCD running coupling following the recipe in
\cite{Larin:1994va,Chetyrkin:1997un,Grozin:2007fh}.}
\begin{align}\label{Ren:constant:Z}
\widetilde{Z}=1+\left(\frac{\alpha_{s}^{\left(n_{l}\right)}\left(\mu_{\Lambda}\right)}{\pi}\right)^{2} \delta \widetilde{Z}^{\left(2\right)}+\left(\frac{\alpha_{s}^{\left(n_{l}\right)}\left(\mu_{\Lambda}\right)}{\pi}\right)^{3}
\delta \widetilde{Z}^{\left(3\right)}+\mathcal{O}(\alpha_s^4),
\end{align}
with
\bea
\delta\widetilde{Z}^{\left(2\right)}=\pi^{2}C_{F}\frac{1}{\epsilon}\left(\frac{3+z}{8\left(1+z\right)}C_{F}+\frac{1}{8}C_{A}\right),\label{eqn:Z2}
\eea
and
\begin{align}\label{eqn:Z3}
& \delta\widetilde{Z}^{\left(3\right)} =\pi^{2}C_{F}\Bigg\{\frac{1}{\epsilon^{2}}\left(\frac{-1+6z}{72\left(1+z\right)}C_{F}^{2}-\frac{5}{48\left(1+z\right)}C_{F}C_{A}-\frac{1}{16}C_{A}^{2}\right)+\frac{1}{\epsilon}\Bigg[\left(\frac{29+38z}{72\left(1+z\right)}-\frac{7}{12}\ln2+\frac{1}{12}\ln\left(1+z\right) \right. \nonumber\\
& \left.-\frac{2-3x-22x^{2}-3x^{3}+2x^{4}}{12\left(1-x\right)\left(1+x\right)^{3}}\ln x+\frac{-1+6z}{24\left(1+z\right)}\ln\frac{\mu_{\Lambda}^{2}}{m_M^2}\right) C_{F}^{2}+\left(\frac{93+52z}{216\left(1+z\right)}+\frac{1}{8}\ln2-\frac{5+2x+5x^{2}}{48\left(1-x\right)\left(1+x\right)}\ln x \right.\nonumber\\
& \left. +\frac{1}{8}\ln\left(1+z\right)+\frac{18+11z}{48\left(1+z\right)}\ln\frac{\mu_{\Lambda}^{2}}{m_M^2}\right)C_{F}C_{A}+\left(\frac{2}{27}+\frac{5}{24}\ln2+\frac{1}{24}\ln\left(1+z\right)+\frac{1}{24}\ln\frac{\mu_{\Lambda}^{2}}{m_M^2}\right)C_{A}^{2}\Bigg]\nonumber\\
& +T_{F}n_{l}\left[\left(\frac{3+z}{36\left(1+z\right)}\frac{1}{\epsilon^{2}}-\frac{15+7z}{108\left(1+z\right)}\frac{1}{\epsilon}\right)C_{F}+\left(\frac{1}{36}\frac{1}{\epsilon^{2}}-\frac{37}{432}\frac{1}{\epsilon}\right)C_{A}\right]+T_{F}n_{b}\left(\frac{1}{15\left(1+1/x\right)^{2}}\frac{1}{\epsilon}\right)C_{F}\nonumber\\
& +T_{F}n_{c}\left(\frac{1}{15\left(1+x\right)^{2}}\frac{1}{\epsilon}\right)C_{F}\Bigg\}.
\end{align}
The expression for $\delta\widetilde{Z}^{\left(3\right)}$ is known for the first time.
A new feature arises that $\delta\widetilde{Z}^{\left(3\right)}$ also explicitly depends on the factorization scale $\mu_\Lambda$.
It is also straightforward to verify that the $\delta\widetilde{Z}$ is indeed symmetric under the exchange $x\leftrightarrow 1/x$.
Reassuringly, taking the $x\to 1$ limit, the factor $\widetilde{Z}$ exactly reproduces the expression of $\widetilde{Z}_p$ in~\cite{Egner:2022jot},
the renormalization constant associated with the pseudo-scalar NRQCD current in the equal quark mass case.

Plugging \eqref{Ren:constant:Z} into \eqref{eqn:gamma}, we then obtain the desired anomalous dimensions affiliated with
the NRQCD operator $\chi^{\dagger}_b\psi_c$ at two and three loop orders:
\begin{subequations}
\label{eqn:anomalous:dimensions}
\begin{align}
& \gamma^{(2)}(x) = -\pi^{2}C_{F}\left(\frac{3+z}{4\left(1+z\right)}C_{F}+\frac{1}{4}C_{A}\right),
\\
& \gamma^{(3)}\left(x, {\mu_\Lambda^2\over m_M^2} \right) = - \pi^{2}C_{F}\Bigg[\left(\frac{29+38z}{24\left(1+z\right)} -\frac{7}{4}\ln2
-\frac{2-3x-22x^{2}-3x^{3}+2x^{4}}{4\left(1-x\right)\left(1+x\right)^{3}}\ln x +\frac{1}{4}\ln\left(1+z\right)
\right.
\nonumber\\
 & \left. +\frac{-1+6z}{8\left(1+z\right)}\ln\frac{\mu_{\Lambda}^{2}}{m_M^2}\right)C_{F}^{2}+\bigg(\frac{93+52z}{72\left(1+z\right)}+\frac{3}{8}\ln2-\frac{5+2x+5x^{2}}{16\left(1-x\right)\left(1+x\right)}\ln x +\frac{3}{8}\ln\left(1+z\right)\nonumber\\
 & +\frac{18+11z}{16\left(1+z\right)}\ln\frac{\mu_{\Lambda}^{2}}{m_M^2}\bigg)C_{F}C_{A}+\left(\frac{2}{9}+\frac{5}{8}\ln2+\frac{1}{8}\ln\left(1+z\right)+\frac{1}{8}\ln\frac{\mu_{\Lambda}^{2}}{m_M^2}\right)C_{A}^{2}\nonumber\\
 & -T_{F}n_{l}\left(\frac{15+7z}{36\left(1+z\right)}C_{F}+\frac{37}{144}C_{A}\right)+T_{F}n_{b}\frac{1}{5\left(1+1/x\right)^{2}}C_{F}+T_{F}n_{c}\frac{1}{5\left(1+x\right)^{2}}C_{F}\Bigg].
\end{align}
\end{subequations}
The two-loop anomalous dimension $\gamma^{(2)}(x)$ was first given in \cite{Onishchenko:2003ui},
later confirmed by \cite{Chen:2015csa}.
The three-loop anomalous dimension $\gamma^{(3)}(x)$ is new, which bears a
a rather complicated form and also explicitly depends on $\ln \mu_\Lambda$.

\begin{figure}[tbh]
  \center{
    \includegraphics*[scale=0.9]{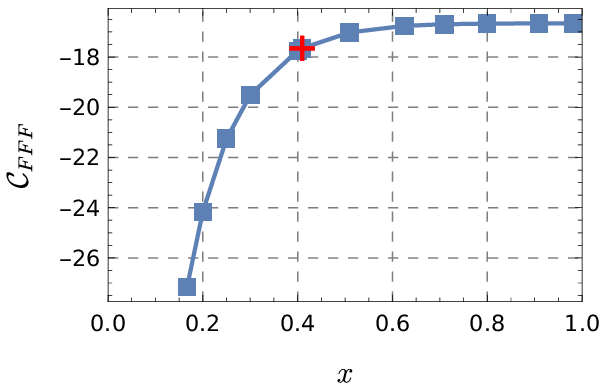}
    \includegraphics*[scale=0.9]{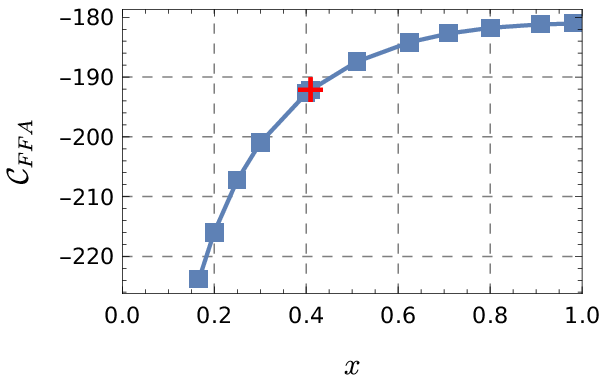}
    \includegraphics*[scale=0.9]{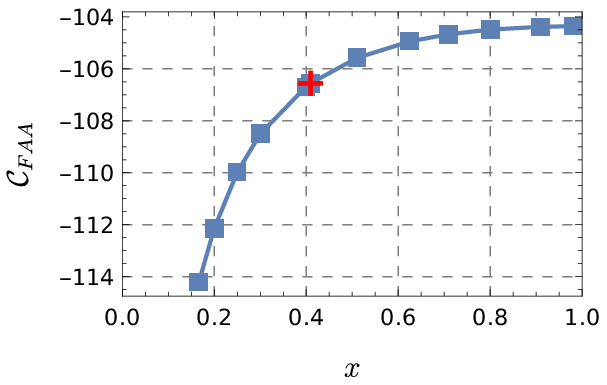}
    \includegraphics*[scale=0.9]{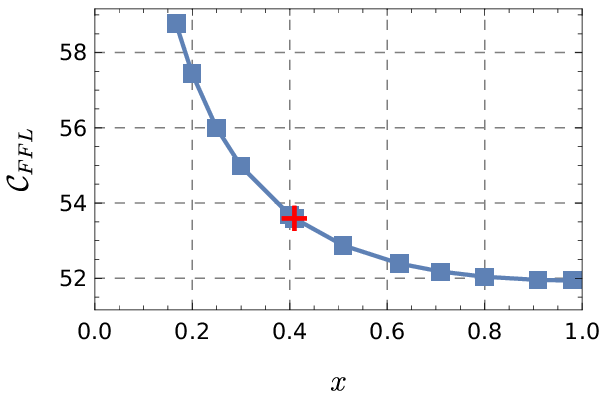}
    \includegraphics*[scale=0.9]{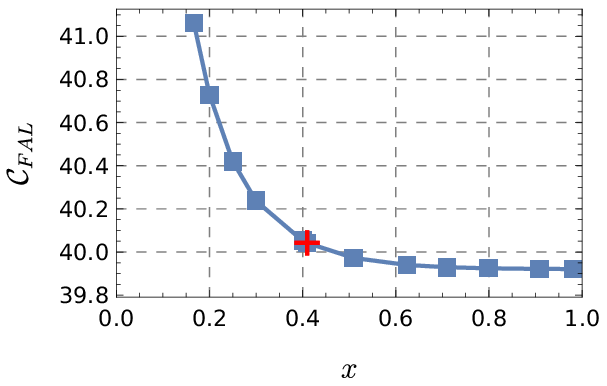}
    \includegraphics*[scale=0.9]{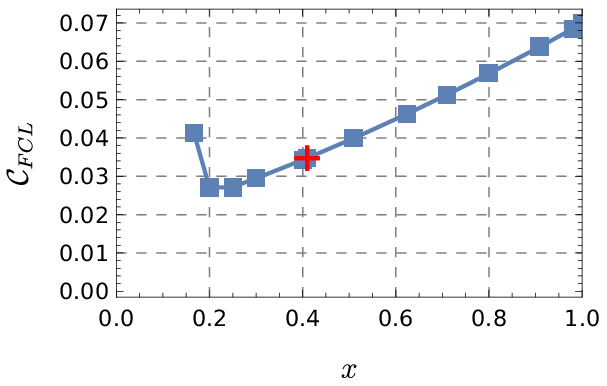}
    \includegraphics*[scale=0.9]{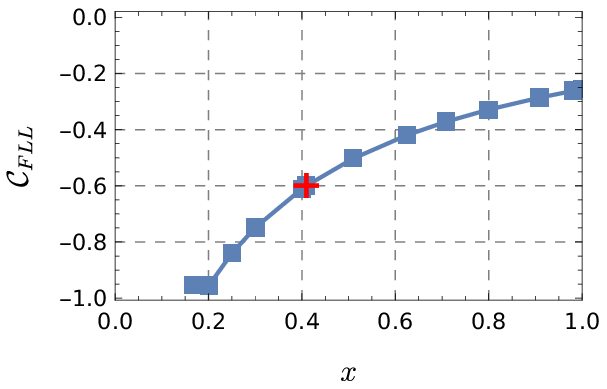}
    \includegraphics*[scale=0.9]{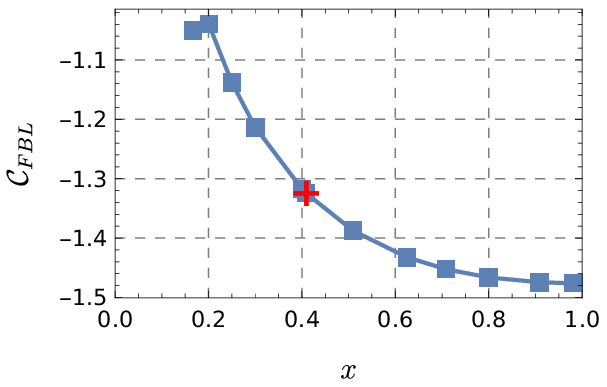}
    \includegraphics*[scale=0.9]{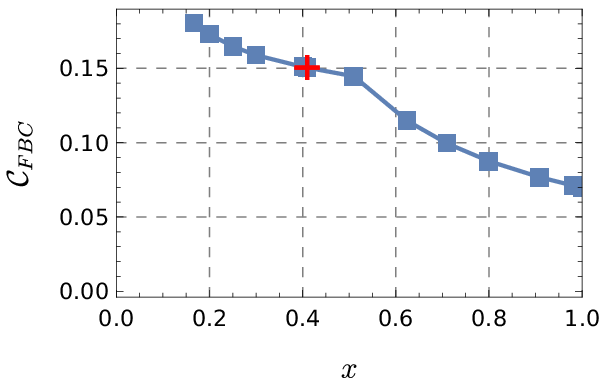}
    \includegraphics*[scale=0.9]{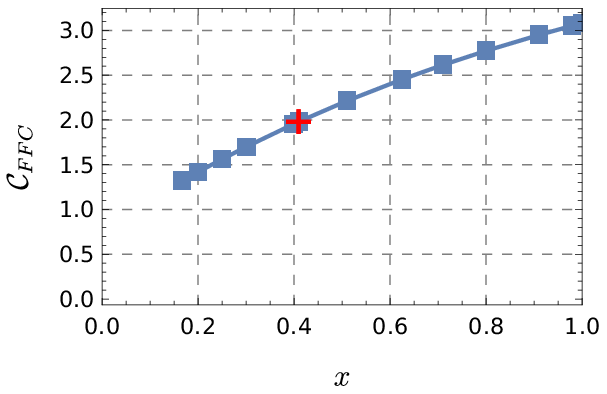}
    \includegraphics*[scale=0.9]{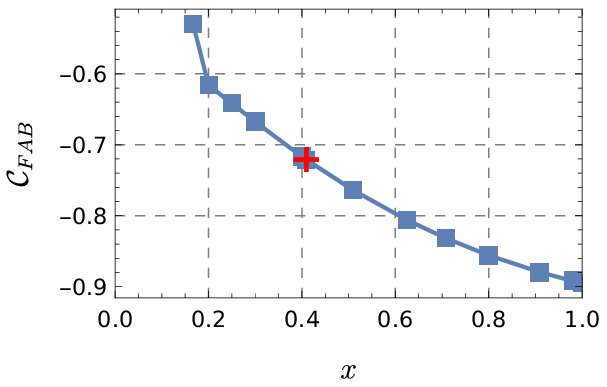}
    \includegraphics*[scale=0.9]{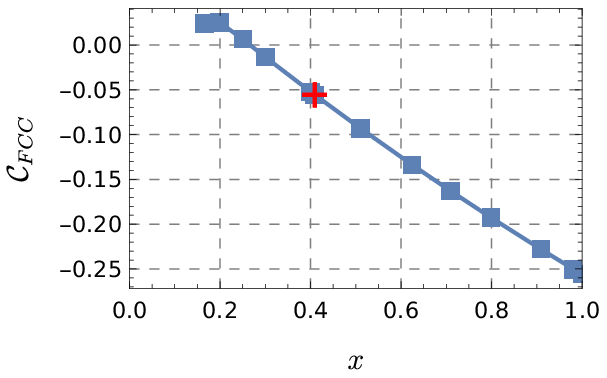}
    \includegraphics*[scale=0.9]{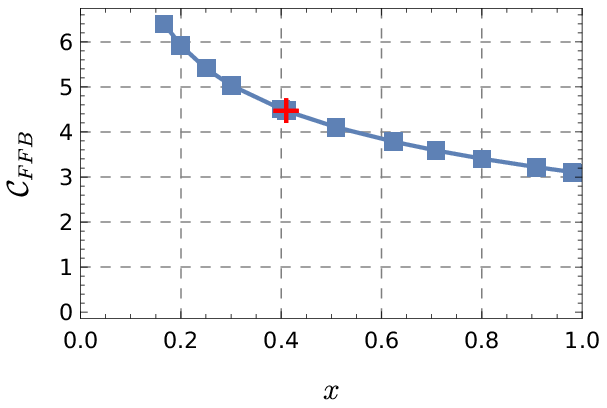}
    \includegraphics*[scale=0.9]{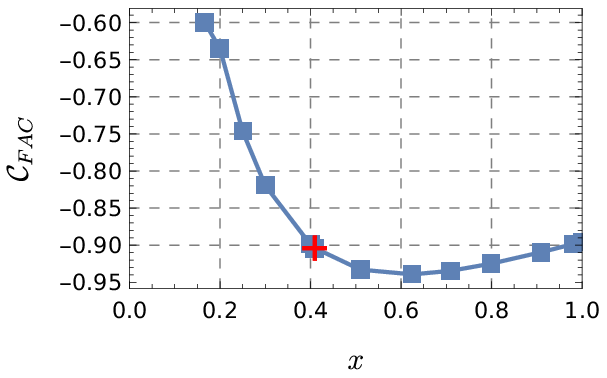}
    \includegraphics*[scale=0.9]{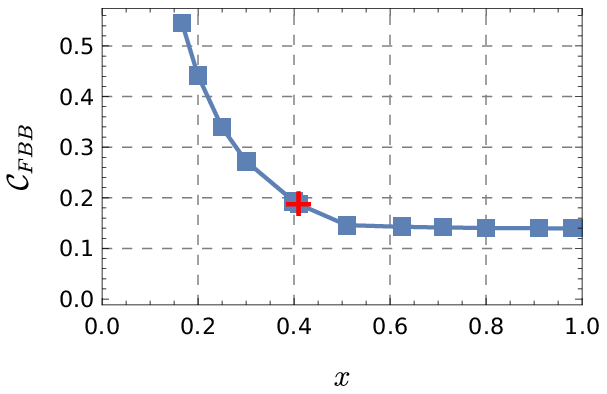}
    \caption {\label{fig:SDCs}
    The profiles of the ${\cal C}^{(3)}(x)$ function according to different color/flavor structure, with $x\in (0,1]$.
    The red cross corresponds to $x_{\rm phys}=2.04/4.98$.
    The rightmost point is evaluated $x=0.98$, which agrees well with the corresponding result for the NRQCD
    pseudoscalar current in equal mass case ($x=1$)~\cite{Egner:2022jot}.
    }
  }
\end{figure}

The only remaining piece in the three-loop SDC in \eqref{eqn:sdc:decomposition} is the term independent of $\ln \mu_R$ and $\ln \mu_\Lambda$,
denoted by $\mathcal{C}^{(3)}(x)$. Following the convention of \cite{Marquard:2014pea,Beneke:2014qea,Egner:2022jot,Feng:2022vvk}, we find it
convenient to decompose the $\mathcal{C}^{(3)}(x)$ in terms of different color/flavor structure:
\begin{align}
\mathcal{C}^{(3)}(x) &=  C_F\Big\{ C^2_F \, \mathcal{C}_{FFF}(x)+C_F \,C_A \, \mathcal{C}_{FFA}(x)
+ C_A^2 \, \mathcal{C}_{FAA}(x) + T_F n_L\, \left[C_F\,\mathcal{C}_{FFL}(x) +C_A\,\mathcal{C}_{FAL}(x)
+ T_F \, n_c \, \mathcal{C}_{FCL}(x) \right.
\nonumber\\
    &  + T_F \, n_b \, \mathcal{C}_{FBL}(x) +T_F \, n_L \, \mathcal{C}_{FLL}(x) \big]
    + T_F^2 \, n_b \, n_c \, \mathcal{C}_{FBC}(x)+
    T_F \, n_c\, \big[ C_F\, \mathcal{C}_{FFC}(x)
    + C_A\, \mathcal{C}_{FAC}(x)+
    T_F \, n_c \, \mathcal{C}_{FCC}(x) \big]
\nonumber\\
&  +T_F\, n_b\, \big[ C_F\, \mathcal{C}_{FFB}(x)+ C_A\,\mathcal{C}_{FAB}(x)+
T_F \, n_b \, \mathcal{C}_{FBB}(x)\big] \Big\}.
\label{eqn:C_col_str}
\end{align}

It is infeasible to obtain the closed functional form for $\mathcal{C}^{(3)}(x)$. Nevertheless, we
are contented with providing highly accurate numerical results. In FIG.~\ref{fig:SDCs} we plot various components
of $\mathcal{C}^{(3)}$ affiliated with each color structure as function of $x$,
Functions with $x>1$ can be mapped by invoking  the $x\leftrightarrow 1/x$ symmetry.

For the forthcoming phenomenological analysis, we start from the precisely known $\overline{\rm MS}$ masses
$\overline{m}_b(\overline{m}_b)= 4.18$ GeV and $\overline{m}_c(\overline{m}_c)=1.28$ GeV~\cite{Workman:2022ynf}.
Using the three-loop formula to convert them into the corresponding pole mass, we obtain $m_{b}=4.98$ GeV $m_c=2.04$ GeV,
with the physical mass ratio $x_{\rm phys} \equiv 2.04/4.98\approx 0.40964$.
Taking this specific reference point, the various components of ${\cal C}^{(3)}$ read:
\begin{subequations}\label{Num:various:C:color:structure}
\begin{align}
&  \mathcal{C}_{FFF}(x_{\rm phys}) = -17.648125254641753539131,
\\
&  \mathcal{C}_{FFA}(x_{\rm phys}) = -192.151798224347908747121,
\\
&  \mathcal{C}_{FAA}(x_{\rm phys}) = -106.55700074027885859242,
\\
&  \mathcal{C}_{FFL}(x_{\rm phys}) = 53.5908823803209988398528,
\\
&  \mathcal{C}_{FAL}(x_{\rm phys}) = 40.041943955625707728391,
\\
&  \mathcal{C}_{FCL}(x_{\rm phys}) = -0.59955659588604920607755,
\\
&  \mathcal{C}_{FBL}(x_{\rm phys}) = -0.05567360504047408860700,
\\
&  \mathcal{C}_{FLL}(x_{\rm phys}) = -1.32484367522413099859707,
\\
 &  \mathcal{C}_{FBC}(x_{\rm phys}) = 0.15047037340977620584792,
\\
&  \mathcal{C}_{FFC}(x_{\rm phys}) = 4.468927007764669701991,
\\
&  \mathcal{C}_{FAC}(x_{\rm phys}) = -0.9039122429495440874057,
\\
&  \mathcal{C}_{FCC}(x_{\rm phys}) = 0.18738217573423910690057,
\\
&  \mathcal{C}_{FFB}(x_{\rm phys}) = 1.9799127987973044694123,
\\
&  \mathcal{C}_{FAB}(x_{\rm phys}) =-0.7210547630289466943049,
\\
&  \mathcal{C}_{FBB}(x_{\rm phys}) = 0.03474911743391490676344.
\end{align}
\end{subequations}

It is curious to assess the impact of this new piece of radiative corrections. Fixing the the renormalization scale $\mu_R$ to be
the reduced quark mass $m_{r}=\frac{m_b m_c}{m_b + m_c}$ (with $m_{r,{\rm phys}}\approx 1.44718$ GeV), and setting the factorization
scale $\mu_\Lambda$ to be $1$ GeV, equation~\eqref{eqn:sdc:decomposition} then reduces to
\begin{align}\label{C:pert:expansion}
    \mathcal{C}(x_{\rm phys}) & =1-1.62623\left(\frac{\alpha_s^{\left(n_l\right)}(m_r)}{\pi}\right)-6.51043\left(\frac{\alpha_s^{\left(n_l\right)}(m_r)}{\pi}\right)^2-1520.59\left(\frac{\alpha_s^{\left(n_l\right)}(m_r)}{\pi}\right)^3+\mathcal{O}(\alpha_s^4)
\end{align}
The ${\cal O}(\alpha_s^3)$ correction looks disquietingly substantial. If taking $\alpha^{(3)}_s(m_r)=0.36406$, the above perturbative series for
the SDC reads $\mathcal{C}(x_{\rm phys})= 1-0.1885-0.08743-2.3663+\mathcal{O}(\alpha_s^4)$. The ${\rm N^3LO}$ correction is even more than twice larger than
the LO result, albeit with opposite sign. Our finding seems to cast some serious doubt on the convergence of perturbative expansion in NRQCD factorization.

We are now ready to make a state-in-the-art prediction to the leptonic decay width of $B_c$:
\begin{align}\label{phenomenological:application}
\Gamma\left(B_c \to l^+ \nu_l\right)
&= \frac{1}{8\pi}\lvert V_{bc}\rvert^2 G_F^2 M_{B_c}  m_l^2\left(1-\frac{m_l^2}{M_{B_c}^2}\right)^2 f_{B_c}^2
\nonumber\\
&= \frac{1}{4\pi}\lvert V_{bc}\rvert^2 G_F^2
 m_l^2 \left(1-\frac{m_l^2}{M_{B_c}^2}\right)^2
 \left\vert \mathcal{C}(x_{\rm phys},\mu_{\Lambda},\mu_{R})\right\vert^2
  \times
    | \langle 0 |\chi^{\dagger}_b\psi_c\left(\mu_{\Lambda}\right) | B_c \rangle|^2
\nonumber\\
&\approx  \frac{1}{4\pi}\lvert V_{bc}\rvert^2 G_F^2
m_l^2 \left(1-\frac{m_l^2}{M_{B_c}^2}\right)^2
\left\vert \mathcal{C}(x_{\rm phys},\mu_{\Lambda},\mu_{R})\right\vert^2 \times
\frac{N_c}{2 \pi} \Big|\overline{R}(\mu_\Lambda)\Big|^2,
\end{align}
where $V_{bc}$ denotes the Cabibbo-Kobayashi-Maskawa (CKM) matrix element,  $m_l$ represents the charged lepton mass,
and $G_F$ denotes the Fermi coupling constant of the weak interaction.
In the second line of \eqref{phenomenological:application}
we have implemented the NRQCD factorization formula for $f_{B_c}$ in
\eqref{eqn:fac}.
In the third line of \eqref{phenomenological:application}, we approximate the NRQCD LDME
by the radial Schr\"odinger wave functions at the origin for $B_c$ in quark potential model.
In Table~\ref{tab:R0} we list some estimations of $|R(0)|^2$ from various theoretical
methods.

\begin{table}[htb]
    \centering
    \begin{tabular}{|c|c|c|c|c|c|c|c|}
\hline
     & pNRQCD & Song-Lin & lattice & Martin & Cornell & Log & B-T \\
     & \cite{Kiselev:2000jc,Kiselev:2003uk}
     & \cite{Song:1986ix,Ikhdair:2005xe}
     & \cite{Colquhoun:2015oha}
     & \cite{Ikhdair:2003ry,Ikhdair:2004hg,Ikhdair:2005xe}
     & \cite{Fulcher:1993sk,Ikhdair:2005xe}
     & \cite{Ikhdair:2003ry,Ikhdair:2004hg,Ikhdair:2005xe}
     & \cite{Eichten:1995ch}\\
     \hline
    $\lvert R(0)\rvert^2$ & 1.588 & 1.54 & 1.539 & 1.495 & 1.413 & 1.28 & 1.642 \\
    \hline
\end{tabular}
\caption{Square of the radial wave function at the origin for $B_c$ (in units of $\rm GeV^3$). }
\label{tab:R0}
\end{table}

\begin{figure}[h]
\center{
\includegraphics*[scale=0.9]{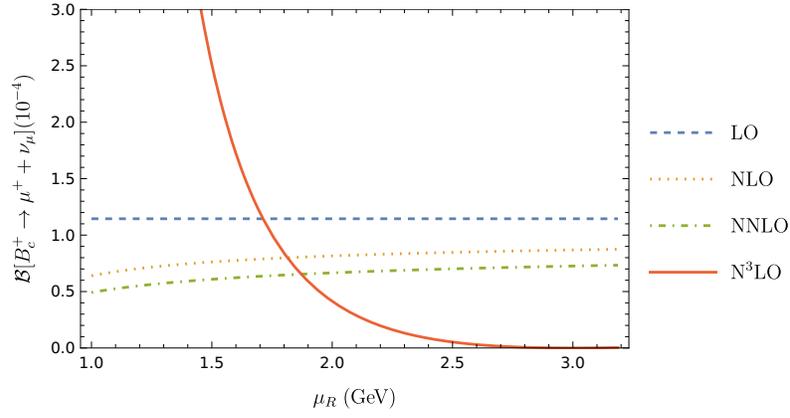}\\
}
\caption {\label{fig:ren} The $\mu_R$ dependence of the predicted branching fractions for $B_c^+\to \mu^+\nu_\mu$ at various perturbative order.
}
\end{figure}

\begin{table}[h]
    \centering
    \begin{tabular}{|c|c|c|c|c|}
\hline
     & LO & NLO & NNLO & $\rm N^3LO$\\
     \hline
     leptonic width($\times 10^{-7}$ eV) & $1.4776$ & $0.97314^{+0.15579}_{-0.14772}$ & $0.77476^{+0.17194}_{-0.13966}$ & $3.9847^{+32.3584}_{-3.9796}$\\
    \hline
   ${\cal B}$($B_c \to \mu^{+} +\nu_{\mu}$) $ (\times 10^{-4})$ & $1.1448$ & $0.75395^{+0.12070}_{-0.11445}$ & $0.60025^{+0.13321}_{-0.10820}$ & $3.0872^{+25.070}_{-3.0832}$\\
    \hline
\end{tabular}
    \caption{The predicted partial widths and the branching ratios for $B_c^+\to \mu^+\nu_\mu$ at various perturbative order. The error is estimated
    by varying $\mu_R$ from $1$ to $3.187$ GeV.}
    \label{tab:br}
\end{table}

In order to make concrete prediction, we fix the values of various input parameters from the latest
\texttt{PDG} compilation~\cite{Workman:2022ynf}:
$V_{bc}=0.0408$, $G_F=1.16638\times 10^{-5}\;\text{GeV}^{-2}$, $M_{B_c}=6.274$ GeV,
$m_{\mu}=0.10566$ GeV.
Note that although both the SDC and the NRQCD LDME logarithmically depend on $\mu_{\Lambda}$,
but their product is independent of $\mu_{\Lambda}$, as guaranteed by the validity of NRQCD factorization theorem.
We choose the central value of the radial wave function at the origin from the NRQCD lattice prediction~\cite{Colquhoun:2015oha},
$|R(0)|^2\vert_{\rm lat}=1.539$ $\rm GeV^3$, which roughly correspond to a scale $\mu_{\Lambda}=1$ GeV respectively.
We evaluate the running QCD coupling with three active flavors using the three-loop formula with the aid of the package \texttt{RunDec}~\cite{Herren:2017osy},
taking the central value $\mu_{R}= m_{r,{\rm phys}} \approx 1.4472 $ GeV.
The theoretical uncertainly is estimated by varying $\mu_R$ from 1 GeV to $m_{M,{\rm phys}}= \sqrt{m_b m_c}\approx 3.1874$ GeV.
Taking the \texttt{PDG} value $\tau(B_c)=0.51 \text{ps}$, we present our predictions to the $B_c$
leptonic width as well as the corresponding branching fraction in Fig.~\ref{fig:ren} and Table~\ref{tab:br}.

We clearly see the $\rm N^3LO$ perturbative correction has overwhelmingly important effect. At first sight, there seems to exist
severe contradiction between \eqref{C:pert:expansion} and Table~\ref{tab:br}, since the ${\rm N^3LO}$ correction in $\cal C$ is deeply negative,
while the ${\rm N^3LO}$ correction significantly enhance the predicted leptonic width. This contradiction arises because when we
square the $\cal C$ in \eqref{phenomenological:application}, we no longer truncate the perturbative series literally up to the order $\alpha_s^{3}$.
But this is a clear sign that alarmingly large three-loop QCD correction has seriously obstructed the perturbative convergence for NRQCD factorization.

Counterintuitively,  from Fig.~\ref{fig:ren} we also see that the predicted branching fraction exhibits rather strong
renormalization scale dependence after incorporating three-loop correction. This can be attributed to the fact that
$\mathcal{C}_{FFA}$ and $\mathcal{C}_{FAA}$ terms in \eqref{Num:various:C:color:structure} turn out to be accidentally large and
negative, which counteract the effect of the explicit $\ln \mu_{R}$ terms in~\eqref{eqn:sdc:decomposition}.

In summary, we have considered the $\rm N^3LO$ QCD corrections to the $B_c$ leptonic decay within the framework of NRQCD factorization.
For the first time, we deduce the analytical expressions of the three-loop renormalization constant of the NRQCD pseudoscalar current, as well
as the corresponding three-loop anomalous dimension associated with $B_c$. Since this anomalous dimension
is a function of the mass ratio between bottom and charm quarks rather than a constant,
the reconstruction of which appears to be much more nontrivial relative to the three-loop QCD corrections to $\Upsilon$ leptonic decay.
Meanwhile, the three-loop short-distance coefficient have also been obtained with exquisite high numerical accuracy.
On the phenomenological perspective, the $\rm N^3LO$ perturbative corrections to $B_c\to l\nu$ is alarmingly huge, and exhibits very
strong dependence on the renormalization scale. In our opinion, our calculation casts some serious doubt on the perturbative convergence
of NRQCD factorization for $B_c$ decay. How to ameliorate this
situation seems to pose some pressing challenge for NRQCD factorization approach.

\begin{acknowledgments}
The work of F. F. is supported by the National
Natural Science Foundation of China under Grant No. 11875318, No. 11505285, and by
the Yue Qi Young Scholar Project in CUMTB.
The work of Y.~J., Z.~M., J.~P and J.-Y.~Z. is supported in part by the National Natural Science Foundation of China under Grants No. 11925506, 11875263,
No. 11621131001 (CRC110 by DFG and NSFC).
The work of W.-L. S. is supported by the National Natural Science Foundation of China
under Grants No. 11975187.
\end{acknowledgments}


\end{document}